\documentstyle[pra,aps,twocolumn,epsf]{revtex}

\begin{document}
\draft
\title{Quantum Teleportation with a Complete Bell State Measurement}
\author{Yoon-Ho Kim,\thanks{Email: yokim@umbc.edu}
Sergei P. Kulik,\thanks{Permanent Address: Department of Physics,
Moscow State University, Moscow, Russia} and Yanhua Shih}
\address{Department of Physics, University of Maryland, Baltimore
County, Baltimore, Maryland 21250}
\date{Submitted to PRL}
\maketitle

\widetext

\vspace*{-13mm}

\begin{abstract}
We report a quantum teleportation experiment in which nonlinear
interactions are used for the Bell state measurements. The
experimental results demonstrate the working principle of
irreversibly teleporting an unknown arbitrary quantum state from
one system to another distant system by disassembling into and
then later reconstructing from purely classical information and
nonclassical EPR correlations. The distinct feature of this
experiment is that {\em all} four Bell states can be distinguished
in the Bell state measurement. Teleportation of a quantum state
can thus occur with certainty in principle.
\end{abstract}


\narrowtext

\vspace*{-5mm}

The idea of quantum teleportation is to utilize the nonlocal correlations
between an Einstein-Podolsky-Rosen pair of particles \cite{EPR} to prepare a
quantum system in some state, which is the exact replica of an arbitrary
unknown state of a distant individual system \cite{Telepub}. Three experiments
in this direction were published recently \cite{Zeilinger,DeMartini,Kimble}.

The following conditions must be satisfied in any claim for
quantum teleportation: (i) the input quantum state, which is
teleported in the experiment must be an {\em arbitrary} state,
(ii) there must be an output quantum state which is an
``instantaneous copy" of the input quantum state, and (iii) the
Bell state measurement (BSM) must be able to distinguish the
complete set of the orthogonal Bell states so that the input
state can be
teleported with certainty. 

In this Letter, we experimentally demonstrate a quantum
teleportation scheme which satisfies all three of the above
conditions. The input state is an {\em arbitrary polarization
state} and the BSM can distinguish {\em all} four orthogonal Bell
states so that the state has a $100\%$ certainty to be teleported
in principle. This is because the BSM is based on nonlinear
interactions which are  {\em necessary} and {\em non-trivial}
physical processes for correlating the input state and the
entangled EPR pair \cite{BSM,klyshko}.

\begin{figure}[hbt] \vspace*{-5mm}\centerline{\epsfxsize=3in\epsffile{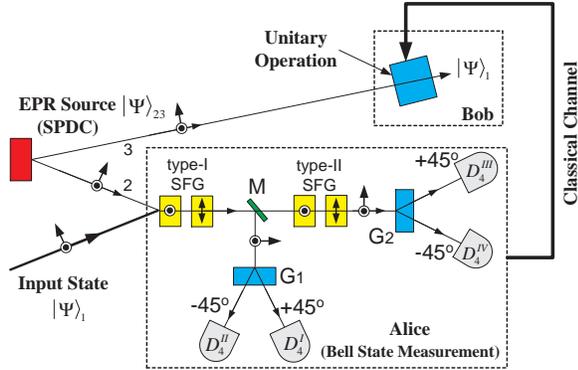}}
\caption{Principle schematic of quantum teleportation with a
complete BSM. Nonlinear interactions (SFG) are used to perform
the BSM. $\odot$ and $\updownarrow$ represent the respective
horizontal and vertical orientations of the optic axes of the
crystals. }\label{fig:schematic}
\end{figure}

The basic elements of the experiment are schematically shown in
Fig. \ref{fig:schematic}. Just as the original proposal of
quantum teleportation \cite{Telepub}, it consists of four
essential parts: (a) the input state, (b) the EPR pair, (c) Alice
(who performs the BSM of the input state and her EPR particle),
and (d) Bob (who carries out unitary operations on his EPR
particle). The input quantum state is an {\em arbitrary
polarization state} given by,

\begin{equation}
|\Psi _{1}\rangle =\alpha |0_{1}\rangle +\beta|1_{1}\rangle,
\label{telestate1}
\end{equation}
 where $|0\rangle$ and $|1\rangle $ represent
the two orthogonal linear  polarization bases (specifically in
this paper) $|H\rangle $ \noindent (horizontal)  and $|V\rangle $
 (vertical) respectively.
$\alpha $ and $\beta $ are two arbitrary complex amplitudes with
respect to the $|0\rangle $ and \newpage \vspace*{20mm} \noindent
$|1\rangle $ bases and they satisfy the condition
$|\alpha|^2+|\beta|^2=1$. The EPR pair shared by Alice and Bob is
prepared by spontaneous parametric down conversion (SPDC) as,
\begin{equation}
|\Psi _{23}\rangle =\frac{1}{\sqrt{2}}\{|0_{2}0_{3}\rangle -|
1_{2}1_{3}\rangle\}, \label{telestate23}
\end{equation}
with the subscripts 2 and 3 as labeled in Fig.\ref{fig:schematic}
\cite{bell}. The complete state of the three particles before
Alice's measurement is then,
\begin{eqnarray}
|\Psi _{123}\rangle =\frac{\alpha }{\sqrt{2}}\{
|0_{1}0_{2}0_{3}\rangle -|0_{1}1_{2}1_{3}\rangle\} \nonumber \\ +
\frac{\beta }{\sqrt{2}}\{ | 1_{1}0_{2}0_{3}\rangle -|
1_{1}1_{2}1_{3}\rangle\}. \label{telestate123}
\end{eqnarray}
The four Bell states which form a complete orthonormal basis for
both particle 1 and particle 2 are usually represented as,
\begin{eqnarray*}
|\Phi_{12}^{(\pm)}\rangle &=&\frac{1}{\sqrt{2}}\{
|0_{1}0_{2}\rangle \pm |1_{1}1_{2}\rangle \},
\\ |\Psi_{12}^{(\pm)}\rangle &=&\frac{1}{\sqrt{2}}\{|0_{1}1_{2}\rangle \pm | 1_{1}0_{2}\rangle
\}.
\end{eqnarray*}
State (\ref{telestate123}) can now be re-written in the following
form based on the above orthonormal Bell states,
\begin{equation}
    \begin{array}{c c c c c c c c c}

| \Psi_{123}\rangle = \frac{1}{2} \{ & & |\Phi _{12}^{(+)}\rangle
&(&\alpha |0_{3}\rangle - \beta |1_{3}\rangle &)& & & \\
&+& |\Phi_{12}^{(-)}\rangle &(&\alpha | 0_{3}\rangle + \beta |
1_{3}\rangle &)& & & \\
&+& |\Psi _{12}^{(+)}\rangle &(&-\alpha | 1_{3}\rangle + \beta |
0_{3}\rangle &)& & &\\
&+& |\Psi _{12}^{(-)}\rangle &(& -\alpha | 1_{3}\rangle - \beta |
0_{3}\rangle &)& &\}.& \label{Bellregroup}

    \end{array}
\end{equation}

To teleport the state of particle 1 to particle 3 reliably, Alice
must be able to distinguish her four Bell states by means of the
BSM performed on particle 1 and her EPR particle (particle 2). She
then tells Bob through a classical channel to perform a
corresponding linear unitary operation on his EPR particle
(particle 3) to obtain an exact replica of the state of particle
1. This completes the process of quantum teleportation.

The distinct feature of the scheme shown in
Fig.\ref{fig:schematic} is that the BSM is based on nonlinear
interactions: optical Sum Frequency Generation (SFG) (or
``upconversion"). Four SFG nonlinear crystals are used for
``measuring" and ``distinguishing" the complete set of the four
Bell states. Photon 1 and photon 2 may interact either in the two
type-I crystals or in the two type-II crystals to generate a
higher frequency photon (labeled as photon 4). The projection
measurements on photon 4 (either at the $45^{\circ }$ or at the
$135^{\circ }$ direction) correspond to the four Bell states of
photon 1 and photon 2,  $ |\Phi_{12}^{(\pm)}\rangle$ and
$|\Psi_{12}^{(\pm)}\rangle$.

Let us now discuss the BSM in detail (see
Fig.\ref{fig:schematic}). The first type-I SFG crystal converts
two $|V\rangle$ polarized photons $|1_{1}1_{2}\rangle$ into a
single horizontal polarized photon $|H_{4}\rangle $. Likewise, the
second type-I SFG crystal converts two $|H\rangle$ polarized
photons $|0_{1}0_{2}\rangle $ into a single vertical polarized
photon $|V_{4}\rangle $. The first and the last terms on the
right-hand side in Eq.(\ref{telestate123}) thus become,
\[
|\Psi_{43}\rangle =\alpha |V_{4}0_{3}\rangle -\beta
|H_{4}1_{3}\rangle.
\]
Dichroic beamsplitter $M$ reflects only SFG photons to the
$45^{\circ}$ polarization projector $G_1$. Two detectors
$D_{4}^{^{I}}$ and $D_{4}^{^{II}}$ are placed at the $45^{\circ
}$ and $135^{\circ }$ output ports of $G_1$ respectively.
Denoting the $45^{\circ }$ and $135^{\circ }$ polarization bases
by $| 45^{\circ }\rangle $ and $| 135^{\circ }\rangle $, the
state $| \Psi _{43}\rangle $ may be re-written as,
\begin{eqnarray}
| \Psi _{43}\rangle =\frac{1}{\sqrt{2}}\{| 45^{\circ }\rangle_{4}
(\alpha | 0_{3}\rangle -\beta | 1_{3}\rangle ) \nonumber
\\ +| 135^{\circ }\rangle_{4} (\alpha | 0_{3}\rangle +\beta |
1_{3}\rangle )\},  \label{telestatefinal}
\end{eqnarray}
which gives,
\begin{eqnarray}
| \Psi \rangle _{3\mid D_{4}^{^{I}} } &=&\alpha |
0_{3}\rangle -\beta | 1_{3}\rangle,  \nonumber \\
| \Psi \rangle _{3\mid D_{4}^{^{II}}} &=&\alpha | 0_{3}\rangle
+\beta | 1_{3}\rangle, \label{telestateport}
\end{eqnarray}
i.e., if detector $D_{4}^{^{I}}$ ($45^{\circ })$ is triggered,
the quantum state of Bob's EPR photon (photon 3) is:
\[
| \Psi _{3}\rangle =\alpha | 0_{3}\rangle -\beta | 1_{3}\rangle,
\]
and, if detector $ D_{4}^{^{II}}$ ($135^{\circ })$ is triggered,
the  quantum state of Bob's photon is:
\[
| \Psi _{3}\rangle =\alpha | 0_{3}\rangle +\beta | 1_{3}\rangle.
\]

As we have analyzed above, the $45^{\circ }$ and the $135^{\circ }$ polarized
type-I SFG components in Eq.(\ref{telestatefinal}) correspond to the
superposition of $|0_{1}0_{2}\rangle $ and $| 1_{1}1_{2}\rangle$  which are the
respective Bell states $|\Phi_{12}^{(+)}\rangle$ and $|\Phi_{12}^{(-)}\rangle$.

Similarly, the other two Bell states are distinguished by the
type-II SFG's. The states $| 0_{1}1_{2}\rangle $ and
$|1_{1}0_{2}\rangle $ are made to interact in the first and the
second type-II SFG crystals respectively to generate a higher
frequency photon with either horizontal (the first type-II SFG)
or vertical (the second type-II SFG) polarization. A $45^{\circ}$
polarization projector $G_2$ is used after the type-II SFG
crystals and two detectors $D_{4}^{^{III}}$ and $D_{4}^{^{IV}}$
are placed at the $45^{\circ}$ and the $135^{\circ }$ output
ports of $G_2$ respectively. On the new bases of $45^{\circ}$ and
$135^{\circ}$ for the SFG photon, the second and the third terms
on the right-hand side in Eq.(\ref{telestate123}) thus become,
\begin{eqnarray}
| \Psi _{43}\rangle =\frac{1}{\sqrt{2}}\{| 45^{\circ }\rangle_{4}
(-\alpha | 1_{3}\rangle +\beta | 0_{3}\rangle ) \nonumber
\\ +| 135^{\circ
}\rangle_{4} (-\alpha | 1_{3}\rangle -\beta | 0_{3}\rangle )\},
\label{telestatefinal2}
\end{eqnarray}
which gives,
\begin{eqnarray}
| \Psi \rangle _{3\mid D_{4}^{^{III}}} &=&-\alpha |
1_{3}\rangle +\beta | 0_{3}\rangle,  \nonumber \\
| \Psi \rangle _{3\mid D_{4}^{^{IV}}} &=&-\alpha | 1_{3}\rangle
-\beta | 0_{3}\rangle, \label{telestateport2}
\end{eqnarray}
i.e., if detector $D_{4}^{^{III}}$ ($45^{\circ })$ is triggered,
the quantum state of Bob's photon is:
\[
| \Psi _{3}\rangle =-\alpha | 1_{3}\rangle +\beta | 0_{3}\rangle,
\]
and if detector $ D_{4}^{^{IV}}$ ($135^{\circ })$ is triggered,
the quantum state of Bob's photon is:
\[
| \Psi _{3}\rangle =-\alpha | 1_{3}\rangle -\beta | 0_{3}\rangle.
\]
The $45^{\circ }$ and the $135^{\circ }$ polarized type-II SFG components
correspond to the superposition of $| 0_{1}1_{2}\rangle $ and $|
1_{1}0_{2}\rangle $ which are the Bell states  $|\Psi _{12}^{(+)}\rangle$ and $
| \Psi _{12}^{(-)}\rangle$ respectively.

To obtain the exact replica of the state of
Eq.(\ref{telestate1}), Bob needs simply to perform a
corresponding unitary transformation after learning from Alice
which of her four detectors, $D_{4}^{^{I}}$, $ D_{4}^{^{II}}$,
$D_{4}^{^{III}}$, or $ D_{4}^{^{IV}}$, has triggered
\cite{bennett}.

\begin{figure}[hbt] \vspace*{-5mm}\centerline{\epsfxsize=3in\epsffile{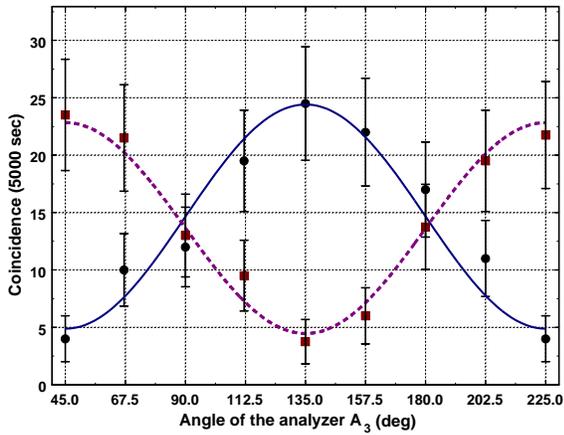}}
\caption{Solid line (circled data points) is the joint detection
rate $D_{4}^{^{I}}$-$D_{3}$ for $45^{\circ }$ linear polarization
as an input state. Dashed line (square data points) is for
$D_{4}^{^{II}}$-$D_{3}$ for the same input state. The expected
$\pi$ phase shift is clearly demonstrated. }\label{fig:typeI}
\end{figure}

\begin{figure}[hbt]
\vspace*{-5mm}\centerline{\epsfxsize=3in\epsffile{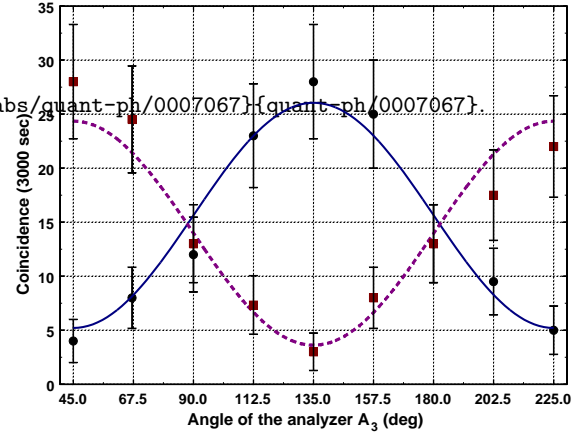}}
\caption{Solid line (circled data points) is the joint detection
rate $D_{4}^{^{III}}$-$D_{3}$ and dashed line (square data
points) is for $D_{4}^{^{IV}}$-$D_{3}$. Again, the expected $\pi$
phase shift is clearly demonstrated. }\label{fig:typeII}
\end{figure}

\begin{figure}[hbt]
\vspace*{-5mm}\centerline{\epsfxsize=3.4in\epsffile{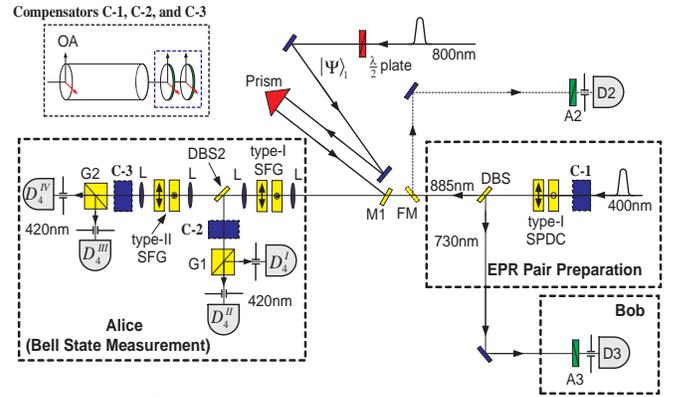}}
\caption{Diagram of the experimental setup. The inset shows the
details of the compensators. See text for
details.}\label{fig:setup}
\end{figure}

To demonstrate the working principle of this scheme, we measure
the joint detection rates between detectors
$D_{4}^{^{I}}$-$D_{3}$, $ D_{4}^{^{II}}$-$D_{3}$,
$D_{4}^{^{III}}$-$D_{3}$ and $ D_{4}^{^{IV}}$-$D_{3}$, where
$D_3$ is Bob's detector (see Fig.\ref{fig:setup}). In these
measurements we choose the input state $|\Psi_{1}\rangle$ as a
linear polarization state. For a fixed input polarization state,
the angle of the polarization analyzer $A_{3}$ which is placed in
front of Bob's detector is rotated and the joint detection rates
are recorded. Figure \ref{fig:typeI} shows two typical data sets
for $D_{4}^{^{I}}$-$D_{3}$ and $ D_{4}^{^{II}}$-$D_{3}$. The input
polarization state is $45^\circ$. Clearly, these data curves
confirm Eq.(\ref{telestateport}). The different phases of the two
curves reflect the phase difference between the two states in
Eq.(\ref{telestateport}). Experimental data for
$D_{4}^{^{III}}$-$D_{3}$ and $ D_{4}^{^{IV}}$-$D_{3}$ show
similar behavior, see Fig. \ref{fig:typeII}, which confirm
Eq.(\ref{telestateport2}).

We now discuss the details of the experimental setup. The
schematic of the experimental setup is shown in Fig.
\ref{fig:setup}. The input polarization state is prepared by
using a $\lambda/2$ plate from a femtosecond laser pulse (pulse
width $\approx$ 100fsec and central wavelength = 800nm)
\cite{note}. The EPR pair (730nm-885nm photon pair) is generated
by two non-degenerate type-I SPDC's. The optical axes of the
first and the second SPDC crystals are oriented in the respective
horizontal ($\odot$) and vertical ($\updownarrow$) directions.
The SPDC crystals are pumped by a $45^\circ$ polarized 100fsec
laser pulse with 400nm central wavelength. The BBO crystals (each
with thickness 3.4mm) are cut for collinear non-degenerate phase
matching. Since the two crystals are pumped equally, the SPDC
pair can be generated either in the first BBO as
$|V_{885}\rangle_2|V_{730}\rangle_3$ ($|1_{2}1_{3}\rangle$) or in
the second BBO as $|H_{885}\rangle_2|H_{730}\rangle_3$ ($|
0_{2}0_{3}\rangle$) with equal probability (885 and 730 refer to
the wavelengths in nanometer). In order to prepare an EPR state
in the form of Eq.(\ref{telestate23}) (a Bell state), these two
amplitudes have to be quantum mechanically ``indistinguishable"
and have the expected relative phase. A Compensator (C-1) is used
for this purpose and it consists of two parts: a thick quartz rod
and two thin plates. The thick quartz rod is used to compensate
the time delay between the two amplitudes $|1_{2}1_{3}\rangle$ and
the $|0_{2}0_{3}\rangle$, and two thin quartz plates are used to
adjust the relative phase between them by angular tilting. A
dichroic beamsplitter $DBS$ is placed behind the SPDC crystals to
separate and send the photon 2 (885nm) and photon 3 (730nm) to
Alice and Bob respectively. To check the EPR state, a flipper
mirror $FM$ is used to send the photon 2 (885nm) to a
photon-counting detector $D_{2}$ for EPR correlation measurement.
Both the space-time and polarization correlations must be checked
before teleportation measurements, in order to be certain of
having high degree EPR entanglement and the expected relative
phase between the $|1_{2}1_{3}\rangle$ and the
$|0_{2}0_{3}\rangle$ amplitudes (see Ref.\cite{twocolor} for
details). Once the EPR state in Eq.(\ref{telestate23}) is
prepared, $FM$ is flipped-down and photon 2 (885nm) is given to
Alice for BSM with photon 1.

The BSM consists of four SFG nonlinear crystals, two $45^{\circ}$
projectors ($G_1$ and $G_2$), four single photon counting
detectors ($D_4^I$, $D_4^{II}$, $D_4^{III}$, and $D_4^{IV}$) and
two compensators as well as other necessary optical components.
The input photon (800nm) and photon 2 (885nm) may either interact
in the two type-I or in the two type-II SFG crystals. Two pairs of
lenses ($L$) are used as telescopes to focus the input beams onto
the crystals. The vertical (horizontal) polarized amplitudes of
the input photon (800nm) and the vertical (horizontal) polarized
photon 2 (885nm) interact in the first (second) type-I SFG to
generate a 420nm horizontal (vertical) polarized photon
\cite{frequency} . The horizontal (vertical) polarized amplitudes
of the input photon and the vertical (horizontal) polarized
photon 2 interact in the first (second) type-II SFG to generate a
420nm horizontal (vertical) polarized photon. The 420nm photons
generated in the type-I SFG process is reflected to detectors
$D_4^I$ and $D_4^{II}$ (after passing through C-2 and a $45^\circ$
polarization projector $G_1$) by a dichroic beamsplitter $DBS_2$
and similarly for the 420nm photons created in two type-II SFG
processes. It is very important to design and adjust the
Compensators (C-2 and C-3) correctly in order to make the
horizontal and the vertical components of the 420nm SFG quantum
mechanically indistinguishable and to attain the expected
relative phase. These two compensators are similar to C-1.

\begin{figure}[hbt] \vspace*{-5mm}\centerline{\epsfxsize=3in\epsffile{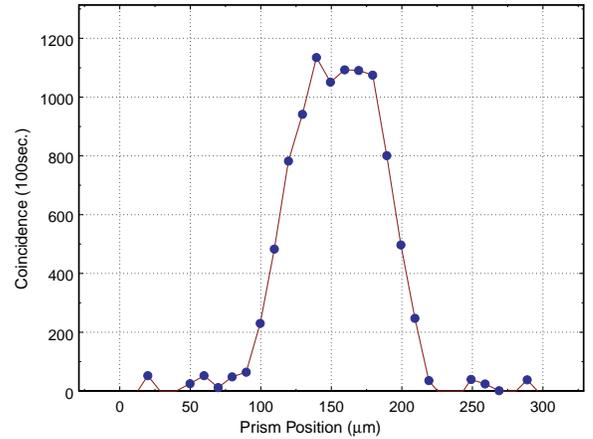}}
\caption{SFG measurement as a function of the prism position. SFG
is observed only when the input pulse and the SPDC photons overlap
exactly inside the crystals. }\label{fig:SFG}
\end{figure}

Since the input state (photon 1) and photon 2 should overlap
inside the SFG crystals exactly, a prism is used to adjust the
path-length of the input pulse \cite{note2}. $M_1$ is a dichroic
mirror which reflects the 800nm photons while transmitting the
885nm ones. In order to be sure that the SFG process occurs with
a single-photon input, we measured the coincidence counting rate
between one of Alice's detectors and Bob's detector $D_{3}$ by
moving the position of the prism. Fig. \ref{fig:SFG} shows a
typical data curve of the measurement. It is clear that SFG only
occurs when the input pulse (photon 1) and photon 2
(single-photon created by the SPDC process) overlap perfectly
inside the SFG crystals \cite{efficiency}.

Readers might have noticed that the efficiency in the
teleportation measurement is a lot lower than the SFG
demonstration. The reason why we get such a low coincidence
counting rate in Figs.\ref{fig:typeI} and \ref{fig:typeII} as
compared to Fig.\ref{fig:SFG} is that very small pinholes had to
be placed in front of Alice's detectors for the teleportation
measurement to ensure good spatial mode overlap. The improvements
of the SFG and the collection efficiencies while preserving good
spatial mode overlap are now underway.

In summary, we have shown a proof-of-principle experimental
demonstration of quantum teleportation with complete a Bell state
measurement. The two main features lie at the heart of our
scheme: (i) EPR-Bohm type quantum correlation and (ii) the BSM
using nonlinear interactions. Single photon SFG is used as the
BSM and the working principle is demonstrated by observing
correlations between the joint measurement of Alice and Bob. In
the current experiment, femtosecond laser pulses are used to
prepare the input polarization state to reduce data collection
time. Recent research on nonlinear optics at low light levels may
enable high-efficiency SFG at single-photon level in the near
future \cite{EIT}.

We would like to thank C.H. Bennett and M.H. Rubin for helpful
discussions. This work was supported in part by the Office of
Naval Research, ARDA, and the National Security Agency.

\begin{references}
\bibitem{bell} Note that any one of the four Bell states can be used for this purpose.
\bibitem{note} In this experiment, the input state is a
polarization state of a femtosecond laser pulse. Only one out of
approximately $10^{10}$ photons, all in the same polarization
state, in each laser pulse is actually ``upconverted" in the SFG
process. It can be easily shown that each photon has the
polarization state of Eq.(\ref{telestate1}) by considering the
correspondence principle. What is being teleported is the state
or qubit associated with this photon. (Note, quantum
teleportation does not teleport the ``quantum" but rather, the
state of the quantum). In principle, it does not prevent one to
use a single-photon qubit as the input state in this experiment.
Due to the low efficiency of SFG, one needs to wait a much longer
time for teleportation to occur.
\bibitem{frequency} Note that there are five different wavelengths which
are tunable as long as the two phase matching conditions (for SPDC
and for SFG) are satisfied. Also five independent relative phases
which affect the output state can be varied.
\bibitem{note2} 800nm input pulse and 400nm pump pulse (which pumps the SPDC crystals) are
actually drawn from a single Ti:Sapphire laser to ensure that
they have the same repetition period.
\bibitem{efficiency} The efficiency of SFG (from the SPDC photons) is roughly estimated to
be 0.1$\sim$1\%. The SFG crystals used for the data shown in this
paper are BBO's with 2mm thickness.
\end{references}
\end{document}